\documentclass{PoS}

\usepackage{amsmath}
\usepackage{slashed}
\usepackage{tabularx}
\usepackage{graphicx}
\usepackage{color}

\newcommand{\I}{\mathrm{i}}
\newcommand{\Tr}{\rm{Tr}}

\newcommand{\separation}{\\[-2mm]}

\graphicspath{{.}{figures/}}

\newcolumntype{L}[1]{>{\raggedright\arraybackslash}p{#1}} 
\newcolumntype{C}[1]{>{\centering\arraybackslash}p{#1}} 
\newcolumntype{R}[1]{>{\raggedleft\arraybackslash}p{#1}} 

\title{Spectral Functions from the Functional Renormalization Group}

\ShortTitle{Spectral Functions from the Functional Renormalization Group}

\newcommand{\TUD}{Theoriezentrum, Institut f\"ur Kernphysik, Technische Universit\"at Darmstadt, Schlossgartenstrasse 2, 64289 Darmstadt, Germany}
\newcommand{\JLU}{Institut f\"ur Theoretische Physik, Justus-Liebig-Universit\"at Giessen, Heinrich-Buff-Ring 16, 35392 Giessen, Germany}
\newcommand{\ECT}{European Centre for Theoretical Studies in Nuclear Physics and Related Areas (ECT*) and Fondazione Bruno Kessler, Strada delle Tabarelle 286, 38123 Villazzano (TN), Italy}
\newcommand{\BNL}{Physics Department, Brookhaven National Laboratory, Upton, NY 11973, USA}
\newcommand{\ITPHD}{Institut f\"ur Theoretische Physik, Ruprecht-Karls-Universit\"at Heidelberg, Philosophenweg 16, 69120 Heidelberg, Germany}

\author{\speaker{Jochen Wambach}\\
	\ECT, and \\ \TUD\\
	E-mail: \email{jwambach@ectstar.eu}}

\author{Christopher Jung\\
	\JLU\\
	E-mail: \email{Christopher.Jung@theo.physik.uni-giessen.de}}

\author{Fabian Rennecke\\
	\BNL, and \\ \ITPHD\\
	E-mail: \email{frennecke@quark.phy.bnl.gov}}

\author{Ralf-Arno Tripolt\\
	\ECT\\
	E-mail: \email{tripolt@ectstar.eu}}

\author{Lorenz von Smekal\\
	\JLU\\
	E-mail: \email{Lorenz.v.Smekal@theo.physik.uni-giessen.de}}

\abstract{We present results for in-medium spectral functions obtained within the Functional Renormalization Group framework. The analytic continuation from imaginary to real time is performed in a well-defined way on the level of the flow equations. Based on this recently developed method, results for the sigma and the pion spectral function for the quark-meson model are shown at finite temperature, finite quark-chemical potential and finite spatial momentum. It is shown how these spectral function become degenreate at high temperatures due to the restoration of chiral symmetry. In addition, results for vector- and axial-vector meson spectral functions are shown using a gauged linear sigma model with quarks. The degeneration of the $\rho$ and the $a_1$ spectral function as well as the behavior of their pole masses is discussed.}

\FullConference{Critical Point and Onset of Deconfinement\\
	7-11 August, 2017\\
	The Wang Center, Stony Brook University, Stony Brook, NY}

\begin{document}
	
\section{Introduction}

The study of matter under extreme conditions, as realized in various cosmological settings and with properties dominated by the strong interaction (QCD) is at the focus of large research efforts. A central question is the modification for the spectral properties of hadrons from the restoration of symmetries that are broken in the vacuum. Experimentally, QCD matter is studied by colliding heavy ions at relativistic bombarding energies. Among the produced particles, photons and lepton pairs have negligible final-state interactions and are thus ideal probes of the entire space-time evolution of the hot and dense fireball. At low invariant masses the measured dilepton spectra are saturated by the light vector-mesons, which directly couple to the electromagnetic field and decay into lepton pairs. In order to interpret such spectra and to extract information on the effects of chiral symmetry restoration at high temperatures and large chemical potentials as well as on possible phase transitions, a realistic theoretical description of in-medium hadrons, in particular of vector mesons, is mandatory.

The stringent description of the spectral properties of hadrons in extremely hot and dense matter poses a major challenge in QCD and is hampered by several difficulties. In first-principles lattice calculations the sign-problem prevents the application of standard Monte-Carlo techniques to the finite-chemical-potential regime. In addition, spectral properties have to be inferred from Euclidean correlation functions by inversion techniques on a finite set of numerical data points which in general is an ill-posed problem.

In this context, the Functional Renormalization Group (FRG) provides a promising alternative \cite{Polonyi:2001se,Bagnuls:2000ae,Gies:2006wv,Berges:2000ew,Pawlowski:2005xe,Litim:2006ag,vonSmekal:2012vx,Pawlowski:2010ht,Delamotte:2007pf}. This non-perturbative continuum framework can be applied at finite quark-chemical potential without further complications and is capable to avoid the analytic continuation problem by solving the flow equations directly in Minkowski space-time. In the following we will use the method proposed in \cite{Tripolt:2013jra,Kamikado:2013sia} while alternative approaches can for example be found in \cite{Floerchinger2012,Pawlowski:2015mia}. In addition, the FRG is also a very powerful tool for the description of phase transitions since it properly takes into account the effects from quantum and thermal fluctuations. 

In these proceedings, we will focus on the effects of chiral symmetry restoration and the in-medium properties of hadrons near the chiral crossover at vanishing chemical potential. After a brief introduction of the FRG in Sec.~\ref{sec:FRG} and of the analytic continuation scheme in Sec.~\ref{sec:analytic_continuation}, we will use the quark-meson model as a low-energy effective theory of QCD and present results for the sigma and the pion spectral function in Sec.~\ref{sec:QM}. In-medium spectral functions of the $\rho$ vector meson and the $a_1$ axial-vector meson are shown in Sec.~\ref{sec:vector} where we use a gauged linear-sigma model including quarks.

\section{Functional Renormalization Group}
\label{sec:FRG}

\begin{figure}[t]
	\centering\includegraphics[width=0.4\textwidth]{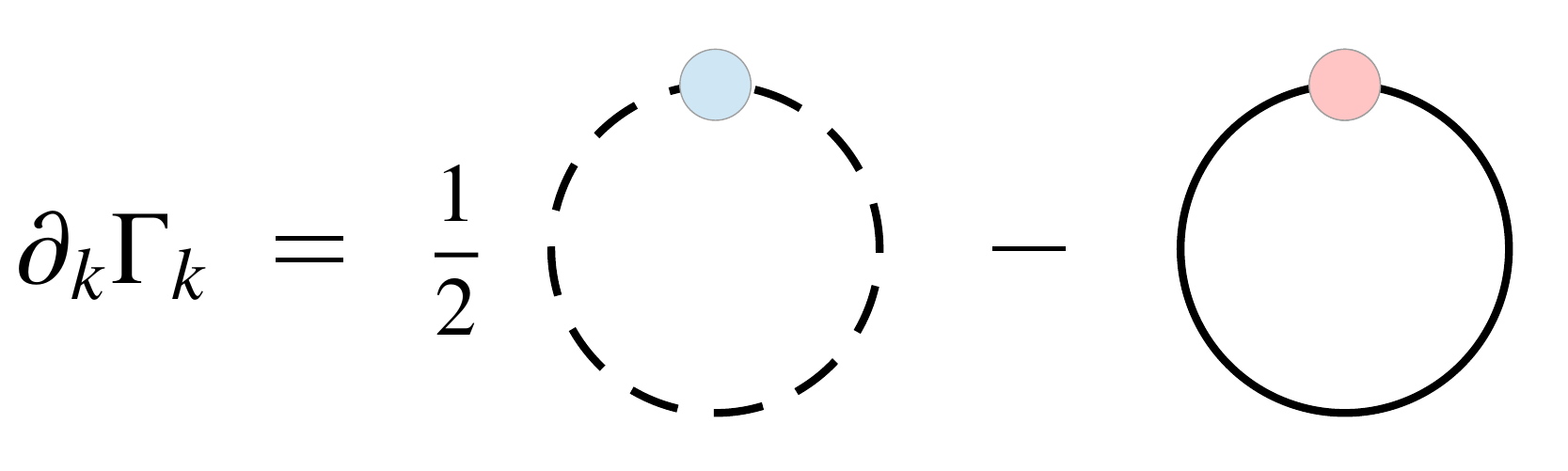}
	\caption{Diagrammatic representation of the flow equation for the effective average action $\Gamma_k$. The dashed line represents a bosonic propagator while the solid line refers to a fermionic propagator. Circles denote derivatives of the regulator functions, $\partial_k R_k$. Figure taken from \cite{Tripolt:2016cey}.}
	\label{fig:flow_gamma_simple} 
\end{figure}

Within the FRG the central object is the resolution scale-dependent effective average action $\Gamma_k$. The RG scale $k$ interpolates between the classical action at some ultraviolet (UV) scale $\Lambda$, $\Gamma_{k=\Lambda}\simeq S$, and the full effective action in the infrared (IR), $\Gamma_{k=0}\equiv \Gamma$. The effective action $\Gamma$ is the generating functional for the one-particle-irreducible (1PI) (Euclidean) correlation functions of a given theory and can be obtained from the generating functional $Z[J]$ for the full Green functions in terms of the following Euclidean path integral over some generic field~$\varphi$,
\begin{align}
Z[J]=\int \mathcal{D}\varphi \:\exp\left( -S[\varphi]+\int d^4x\: J(x)\varphi (x)\right),
\end{align}
where $S[\varphi]$ is the classical action. The generating functional for the connected $n$-point Green functions is then given by
\begin{align}
W[J]=\log Z[J],
\end{align}
and the effective action $\Gamma[\phi]$ can be expressed as the Legendre transform of $W[J]$ with respect to the expectation value $\phi(x) = \langle \varphi(x) \rangle_J $ of the field  $\varphi(x)$ in presence of the source $J(x)$,
\begin{align}
\Gamma[\phi]=\sup_J \left( \int d^4x \:J(x) \phi(x)-W[J]\right).
\end{align}
When evaluated for a homogeneous and space-time independent equilibrium state $\phi= \phi_0$ which satisfies
\begin{align}
\left.\frac{\delta\Gamma[\phi]}{\delta\phi}\right|_{\phi=\phi_0}=0,
\end{align}
the effective action is related to the grand-canonical potential by
\begin{align}
\label{EffTP}
\Omega(T,\mu)=\frac{T}{V}\Gamma[\phi_0].
\end{align}
Intuitively, an RG-scale dependence of $Z[J]$ can be introduced by separating the low-momentum modes from the high-momentum modes:
\begin{align}
\varphi(x)=\varphi_{q\leq k}(x)+\varphi_{q>k}(x),
\end{align}
which yields
\begin{align}
Z[J]=\int \mathcal{D}\varphi_{q\leq k}
\underbrace{
\int \mathcal{D}\varphi_{q>k}
\:\exp\left( -S[\varphi]+\int d^4x\: J(x)\varphi (x)\right)
}_{=Z_k[J]} .
\end{align}
The scale-dependent generating functional $Z_k[J]$ is connected to the full functional $Z[J]$ that contains the effects from all momentum modes by $\lim_{k\rightarrow 0} Z_k[J]=Z[J]$. This reflects Wilson's coarse-graining idea and can be made more explicit by introducing a regulator function $R_k$. The scale-dependent analogues of the generating functionals $W[J]$ and $\Gamma[\phi]$ are then given by
\begin{align}
W_k[J]=\log \int \mathcal{D}\varphi\: \exp \left( -S[\varphi] -\Delta S_k[\varphi]+\int d^4x\: J(x)\varphi(x)\right),\\
\Gamma_k[\phi]=\sup_J\left( \int d^4x\: J(x)\phi(x)-W_k[J]\right) -\Delta S_k[\phi],
\end{align}
where the regulator insertion,
\begin{align}
\Delta S_k[\phi]=\frac{1}{2}\int\frac{d^4q}{(2\pi)^4}\:\phi(-q)R_k(q)\phi(q),
\end{align}
acts as a scale-dependent mass term and suppresses fluctuations of modes with momenta lower than the RG scale $k$. The scale-dependence of the effective average action $\Gamma_k$ is described by the Wetterich equation \cite{Wetterich:1992yh, Morris:1993qb},
\begin{align}
\label{eq:Wetterich}
\partial_k \Gamma_k[\phi]=
\:\,\frac{1}{2}\,\Tr \left\{\partial_kR_k \left(\Gamma_k^{(2)}[\phi]+R_k\right)^{-1}\right\},
\end{align}
where $\Gamma_k^{(2)}[\phi]$ is the second functional derivative w.r.t.~the field $\phi$, and the trace includes an integration over the loop momentum as well as a summation over all internal indices. The Wetterich equation has a simple one-loop representation as shown in Fig.~\ref{fig:flow_gamma_simple}. 

Although the Wetterich equation in principle is exact, truncations are necessary to solve it in practice. A common truncation scheme is given by the so-called `derivative expansion' for $\Gamma_k$, i.e.
\begin{align}
\label{eq:derivative_expansion}
\Gamma_k[\phi]=\int d^4x\left\{U_k(\phi)+\frac{1}{2}Z_k(\phi)(\partial_\mu \phi)^2+\frac{1}{8}Y_k(\phi) (\partial_\mu \phi^2)^2+\mathcal{O}(\partial^4)\right\},
\end{align}
were $U_k(\phi)$ is the effective potential while $Z_k(\phi)$ and $Y_k(\phi)$ are field and scale-dependent derivative terms. Using only the effective potential and neglecting higher-order corrections, called the `local potential approximation' (LPA), already gives a satisfactory quantitative description of the thermodynamics and critical phenomena as obtained in several cases (see, e.g., \cite{Litim:2001dt,Braun:2009si}). 

\section{Analytic continuation of flow equations}
\label{sec:analytic_continuation}

Since the FRG is a Euclidean framework an analytic continuation to Minkowski space-time is necessary for obtaining real-time quantities such as spectral functions. In the following we will use the method proposed in \cite{Tripolt:2013jra,Kamikado:2013sia} which involves a two-step analytic continuation procedure on the level of the flow equations for the two-point functions. First, the Euclidean flow equations are obtained by taking two functional derivatives of the Wetterich equation w.r.t.~the fields, which yields the following generic structure, see also Fig.~\ref{fig:flow_equations_gamma2},
\begin{align}
\partial_k\Gamma_k^{(2)}
&=
\frac{1}{2}\frac{\delta^2}{\delta \phi^2}
\Tr \left\{\partial_kR_k D_k\right\}
\nonumber\\
&=-\frac{1}{2} \Tr 
\left\{
\partial_kR_k \frac{\delta}{\delta \phi}
\left(
D_k\Gamma_k^{(3)}D_k
\right)
\right\}\nonumber\\
&=
\Tr 
\left\{
\partial_kR_k
\left(
D_k\Gamma_k^{(3)}D_k\Gamma_k^{(3)}D_k
\right)
\right\}
-\frac{1}{2}
\Tr 
\left\{
\partial_kR_k\left(
D_k\Gamma_k^{(4)}D_k\right)
\right\},
\end{align}
where $D_k(q)$ is the scale-dependent propagator
\begin{align}
D_k(q)&\equiv\left(\Gamma_k^{(2)}(q)+R_k(q)\right)^{-1},
\end{align}
and $\Gamma_k^{(3)}$ and $\Gamma_k^{(4)}$ are the scale-dependent three- and four point functions. The flow equations for the two-point function depend on the external four-momentum $(p_0,\vec{p})$ while the internal momenta are integrated over. At finite temperature, the integration over the internal energy $q_0$ turns into a Matsubara sum over discrete Euclidean momentum modes. In order to perform the analytic continuation procedure proposed in \cite{Tripolt:2013jra,Kamikado:2013sia}, this Matsubara sum has to be performed analytically, which involves the use of three-dimensional regulator functions and leads to the appearance of bosonic and fermionic occupation number factors, $n_B$ and $n_F$. In a second step the flow equations are analytically continued to real external energies by making use of the periodicity of the occupation numbers w.r.t.~the discrete Euclidean energy $p_0$,
\begin{align}
n_{B,F}(E+i p_0)\rightarrow n_{B,F}(E),
\end{align}
and then replacing $p_0$ by a continuous real energy $\omega$,
\begin{align}
\partial_k\Gamma^{(2),R}_k(\omega,\vec p)=-\lim_{\epsilon\to 0} \partial_k\Gamma^{(2),E}_k(p_0=-\I(\omega+\I\epsilon), \vec p),
\end{align}
where the limit $\epsilon\to 0$ can be taken analytically for the imaginary part. The spectral function is finally obtained as
\begin{align}
\rho(\omega,\vec p)=-\frac{1}{\pi}\text{Im}\frac{1}{\Gamma^{(2),R}_{k\to 0}(\omega,\vec p)}.
\end{align}

\section{Spectral Functions for the Quark-Meson Model}
\label{sec:QM}

\begin{figure}[t]
	\includegraphics[width=0.5\textwidth]{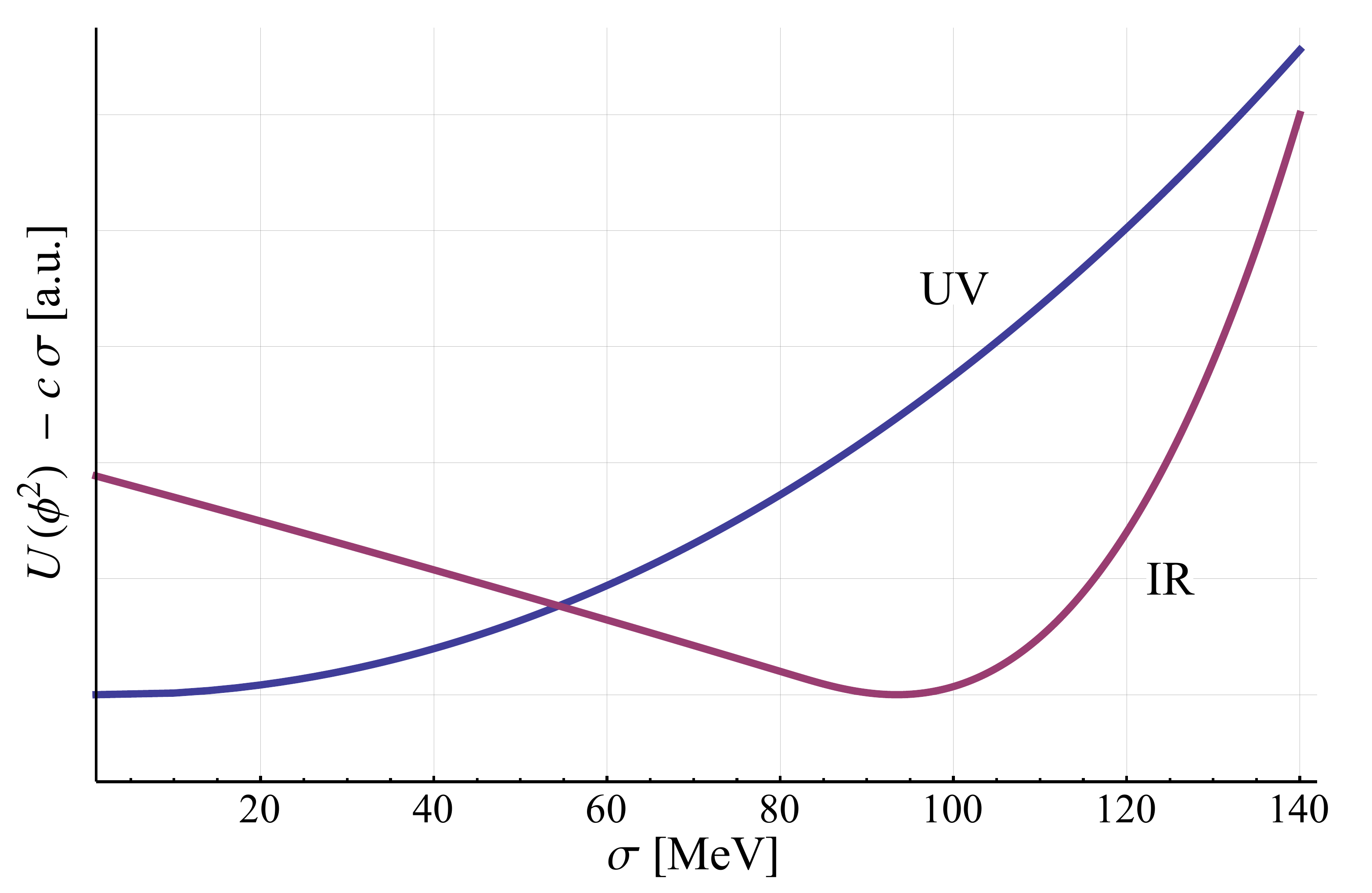}\hspace{5mm}
	\includegraphics[width=0.41\textwidth]{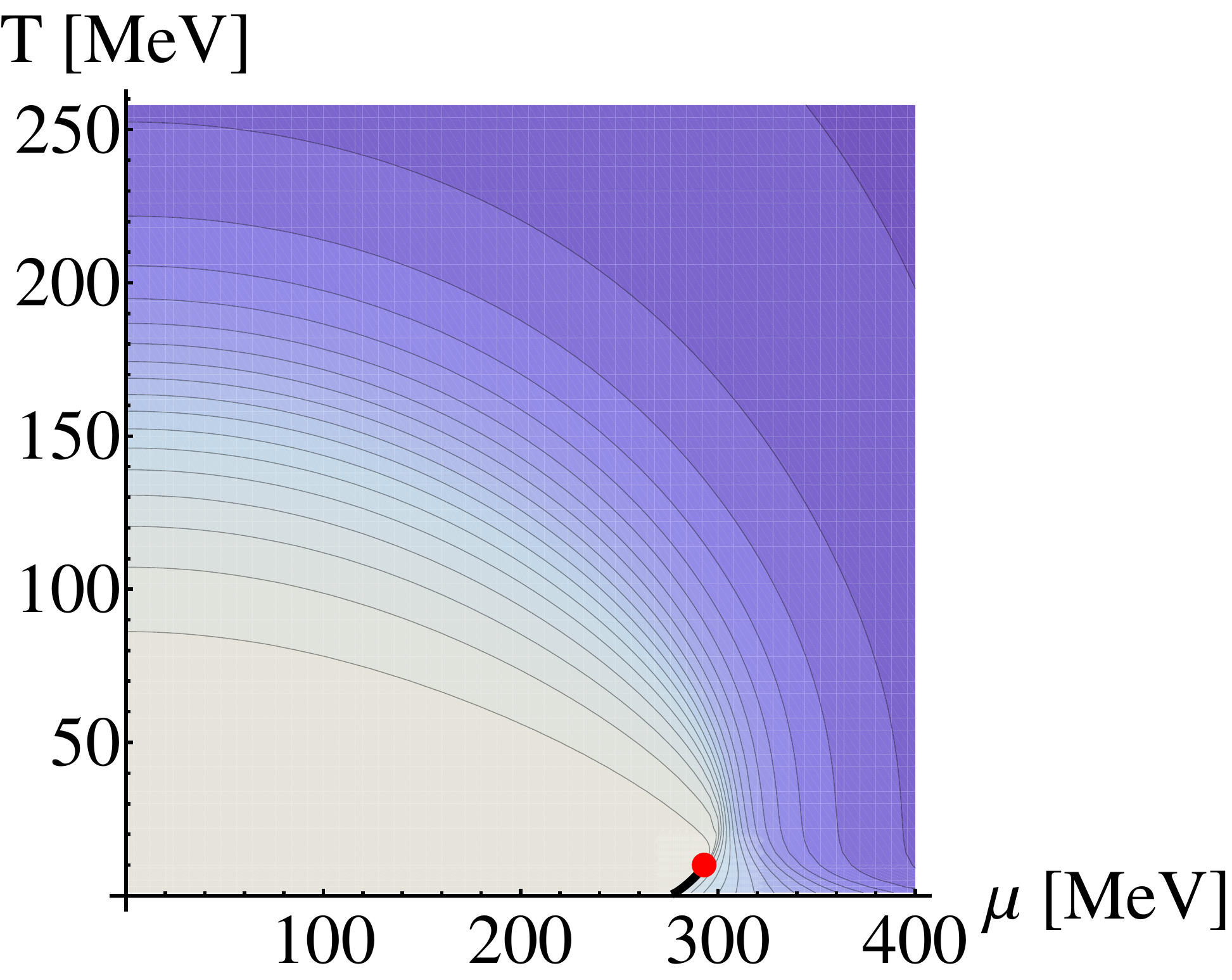}\llap{\makebox[2.2cm][l]{\raisebox{1.5cm}{\includegraphics[height=3.0cm]{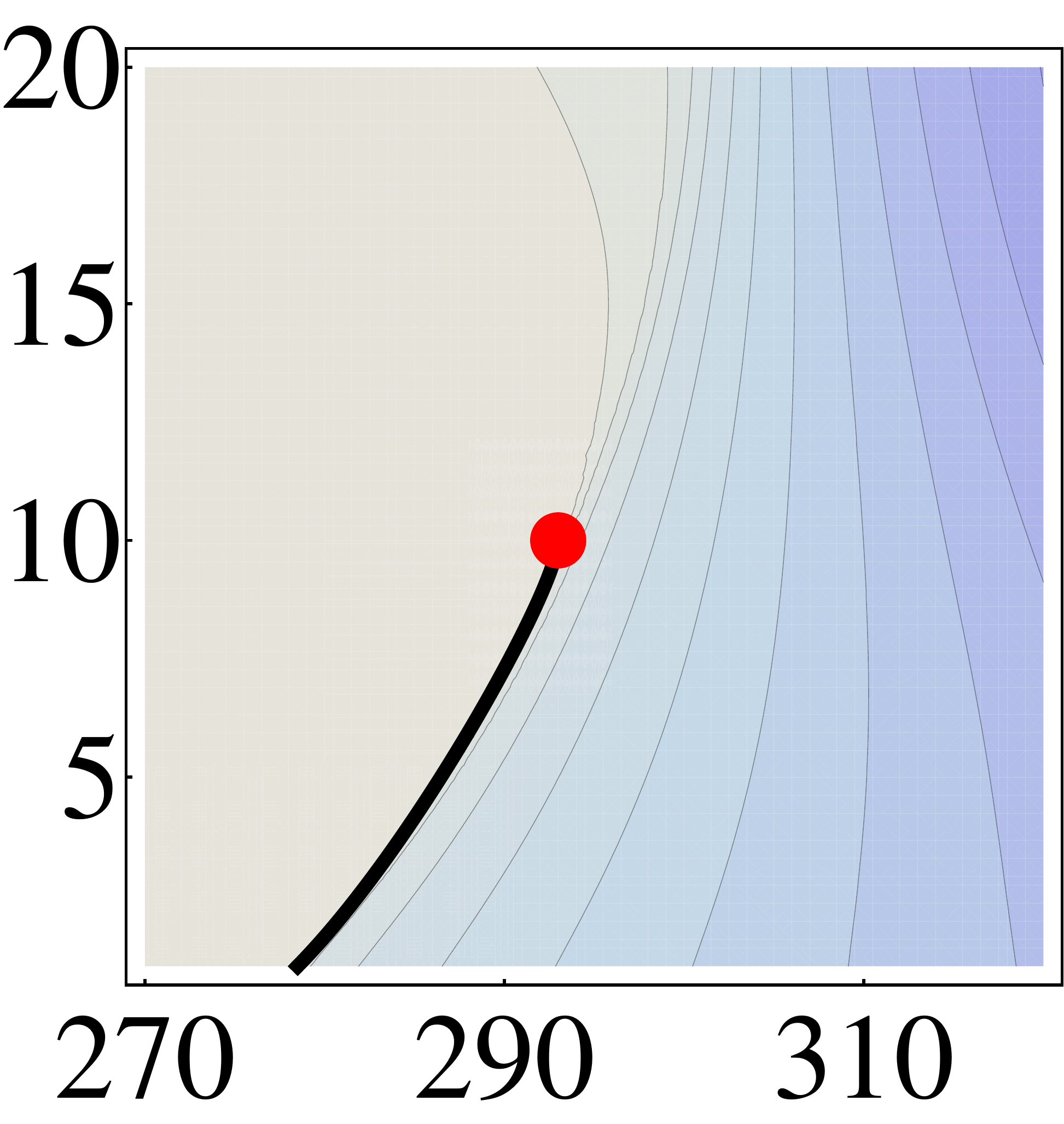}}}}
	\caption{Left: The explicitly broken effective potential, $U_k(\phi^2)-c\sigma$, is shown at the UV scale $k=\Lambda$, as given by Eq.~(\ref{eq:pot_UV}), and in the IR, as obtained by solving the corresponding flow equation in the vacuum. In the IR, chiral symmetry is spontaneously broken and the global minimum of the potential is located at $\sigma_0=f_\pi=93$~MeV. Figure taken from \cite{Tripolt:2016cey}. Right: The phase diagram of the quark-meson model is shown as a contour plot of the order parameter for chiral symmetry as a function of temperature $T$ and quark-chemical potential $\mu$, $\sigma_0(\mu,T)$. In the vacuum, chiral symmetry is spontaneously broken and we have $\sigma_0=f_\pi=93$~MeV, while $\sigma_0$ decreases towards higher $T$ and $\mu$ as chiral symmetry is progressively restored. The inset shows the first-order phase transition and the critical endpoint.}
	\label{fig:pot_and_pd} 
\end{figure}

\begin{figure}[t]
	\includegraphics[width=0.99\textwidth]{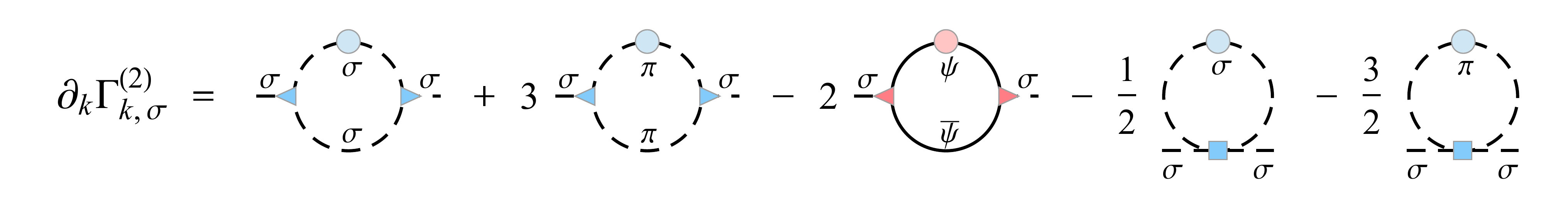}\\
	\includegraphics[width=0.99\textwidth]{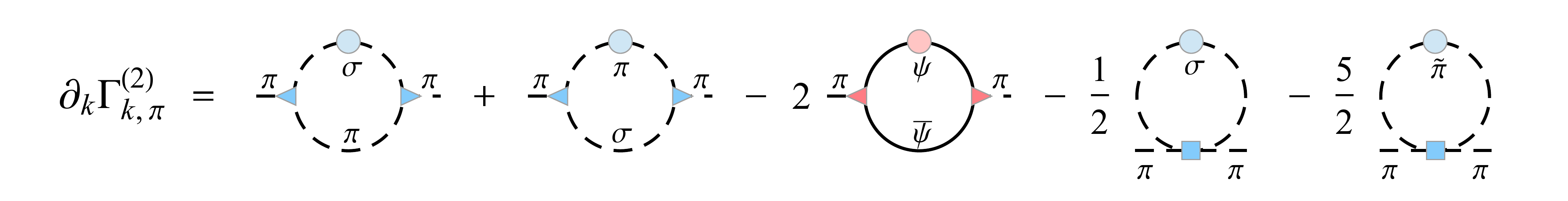}
	\caption{Diagrammatic representation of the flow equations for the sigma and pion two-point functions of the quark-meson model. Dashed lines denote mesonic propagators, solid lines fermionic propagators, circles correspond to regulator insertions, $\partial_k R_k$, while triangles and squares denote three- and four- point vertices, respectively. Figure taken from \cite{Tripolt:2016cey}.}
	\label{fig:flow_equations_gamma2} 
\end{figure}

Concentrating on chiral symmetry aspects we will use the quark-meson (QM) model as a low-energy effective theory for QCD.  To lowest order in the derivative expansion this implies the following Ansatz for the effective average action:
\begin{align}
\label{eq:gamma}
\Gamma_{k}=
\int d^{4}x \:\left\{
\bar{\psi}\left(\gamma_\mu\partial^\mu+
h(\sigma+i\vec{\tau}\vec{\pi}\gamma^{5}) -\mu \gamma_0 \right)\psi
+\frac{1}{2} (\partial_{\mu}\phi)^{2}+U_{k}(\phi^2)-c\sigma
\right\},
\end{align}
where $\phi=(\vec{\pi},\sigma)^T$. At the UV scale $\Lambda$, the effective potential is chosen to be symmetric, i.e. in the chirally unbroken phase
\begin{align}
\label{eq:pot_UV} 
U_\Lambda(\phi^{2}) =
\frac{1}{2}m_\Lambda^{2}\phi^{2} +
\frac{1}{4}\lambda_\Lambda(\phi^{2})^{2},
\end{align}

Explicit values for the parameters are listed in Tab.~\ref{tab:parameters1}. By inserting the Ansatz for $\Gamma_{k}$ into the Wetterich equation, one obtains the flow equation for the effective potential. For explicit expressions of the flow equations, the choice of regulator functions and for details on the numerical implementation we refer to \cite{Tripolt:2013jra}. In Fig.~\ref{fig:pot_and_pd} the shape of the effective potential is shown in the UV and in the IR. On account of the quarks chiral symmetry is spontaneously broken during the scale evolution. The global minimum of the effective potential, $\sigma_0$, acts as the chiral order parameter and is used in Fig.~\ref{fig:pot_and_pd} to obtain the phase diagram. It shows a chiral crossover transition at $T\approx 170$~MeV for $\mu=0$ which turns into a chiral critical endpoint (CEP) at $T_c\approx 9$~MeV and $\mu\approx 292$~MeV and a first-order phase transition for $T<T_c$. The unusual behavior of the first-order line gives rise to a phase with negative entropy as pointed out in \cite{Tripolt:2017zgc}.

\begin{table}[b!]
	\centering
	\begin{tabular}{C{1.2cm}|C{0.6cm}|C{1.5cm}|C{0.6cm}|C{1.7cm}|C{1.5cm}|C{1.5cm}|C{1.5cm}}
		$m_\Lambda/\Lambda$ & $\lambda_\Lambda$ & $c/\Lambda^3$ &  $h$ & $\sigma_0\equiv f_\pi$ & $m_\pi$ & $m_\sigma$ &  $m_\psi$   \\
		\hline\hline
		0.794  & 2 & 0.00175 & 3.2 &93.5 MeV & 138 MeV & 509 MeV & 299 MeV
	\end{tabular}
	\caption{Parameter set used for the quark-meson model and corresponding vacuum values obtained in the IR for the pion decay constant, $f_\pi$, and the Euclidean masses. The UV cutoff is chosen to be $\Lambda=1000$~MeV.}
	\label{tab:parameters1} 
\end{table}

\begin{figure}[t!]
	\centering\includegraphics[width=0.6\textwidth]{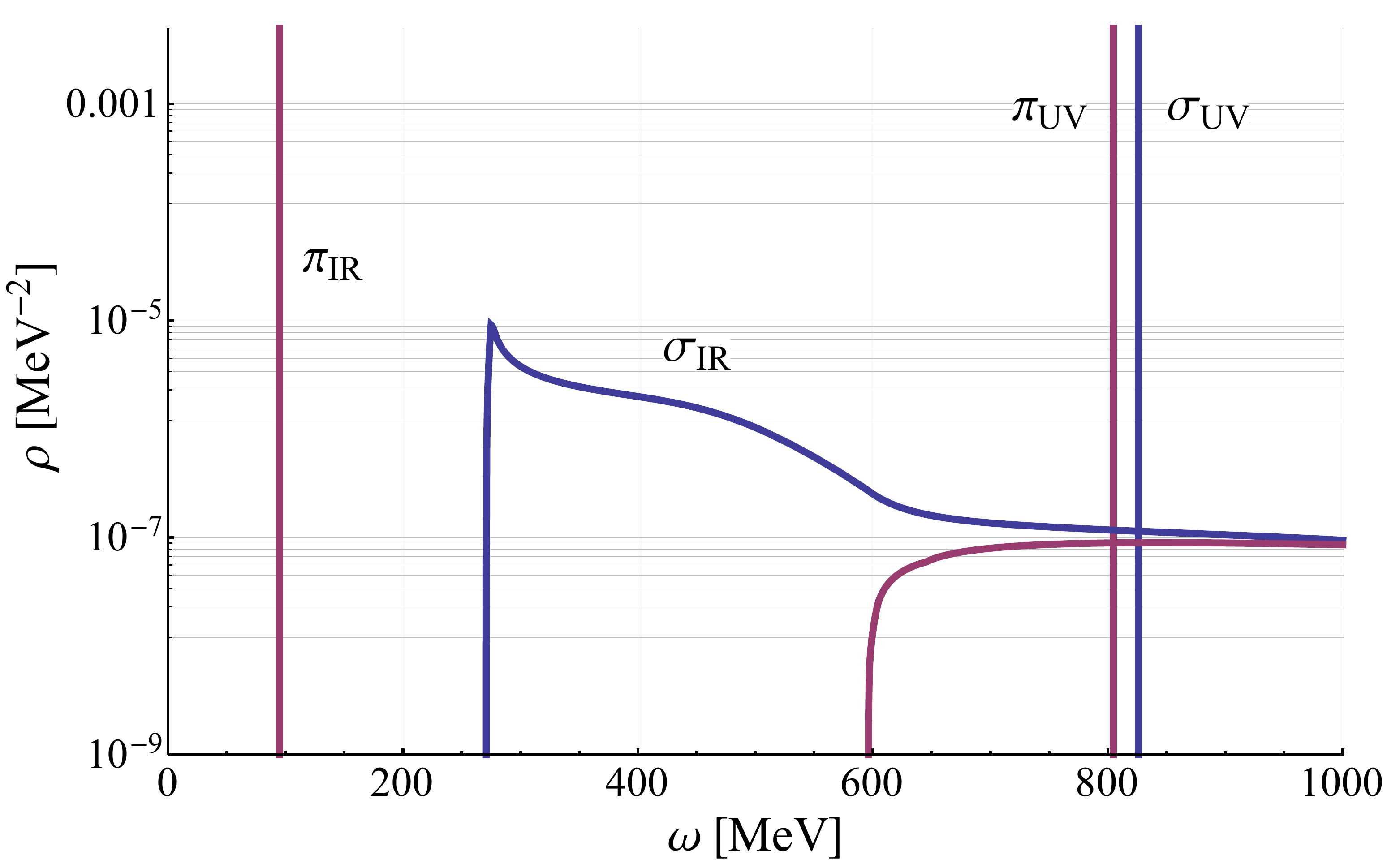}
	\caption{The sigma and pion spectral functions are shown in the vacuum at the UV scale, $k=\Lambda$, and in the IR, at $k=0$. In the UV, the sigma and the pion are stable particles and their spectral functions are given by Dirac delta functions. In the IR, the effects from decay channels have been included which gives rise to a complicated structure of the spectral functions, see text for details. Figure taken from \cite{Tripolt:2016cey}.}
	\label{fig:spectral_UV_IR} 
\end{figure}

\begin{figure}[b!]
	\includegraphics[width=0.5\textwidth]{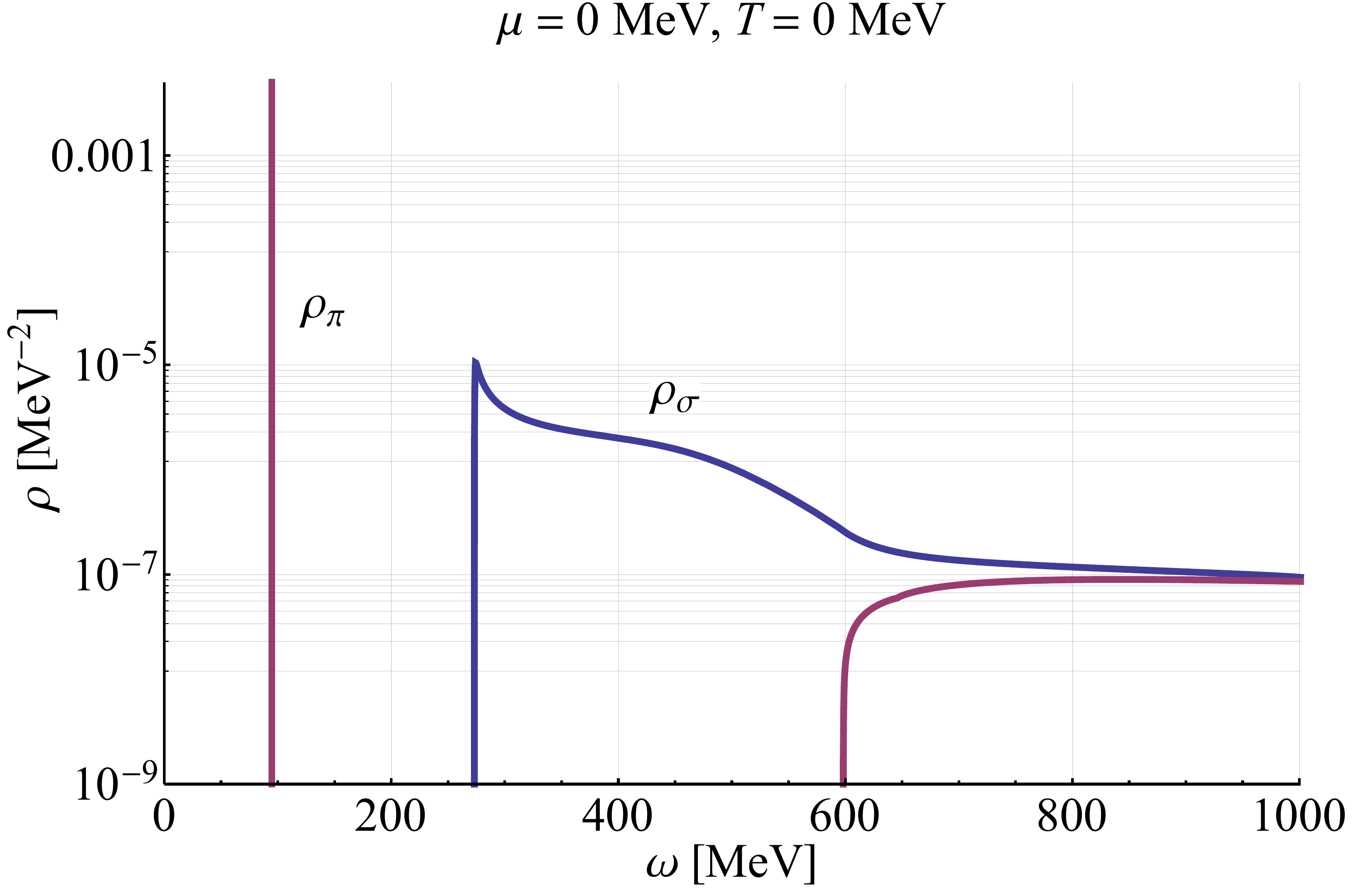}\hspace{2mm}
	\includegraphics[width=0.5\textwidth]{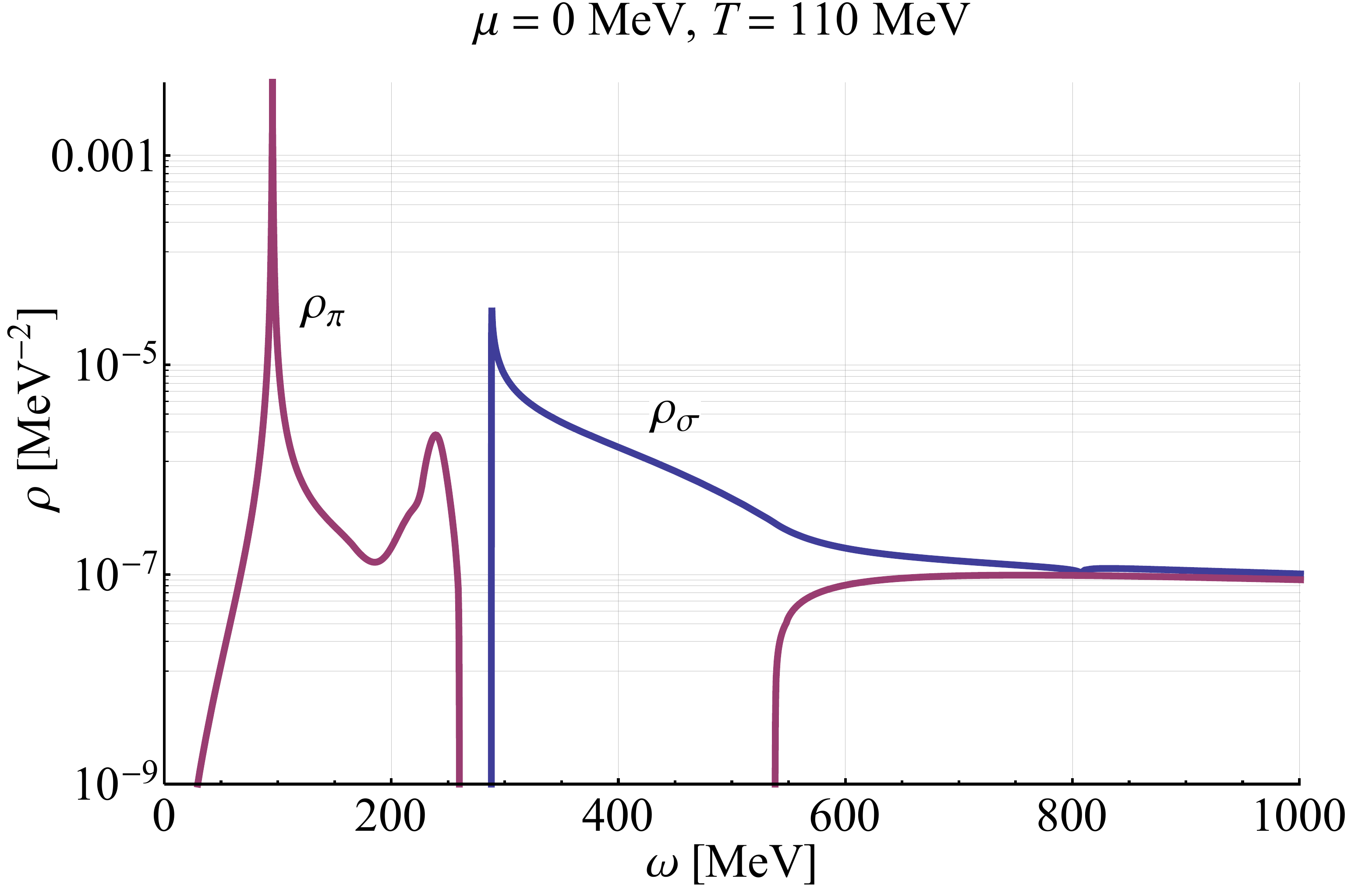}\\[5mm]
	\includegraphics[width=0.5\textwidth]{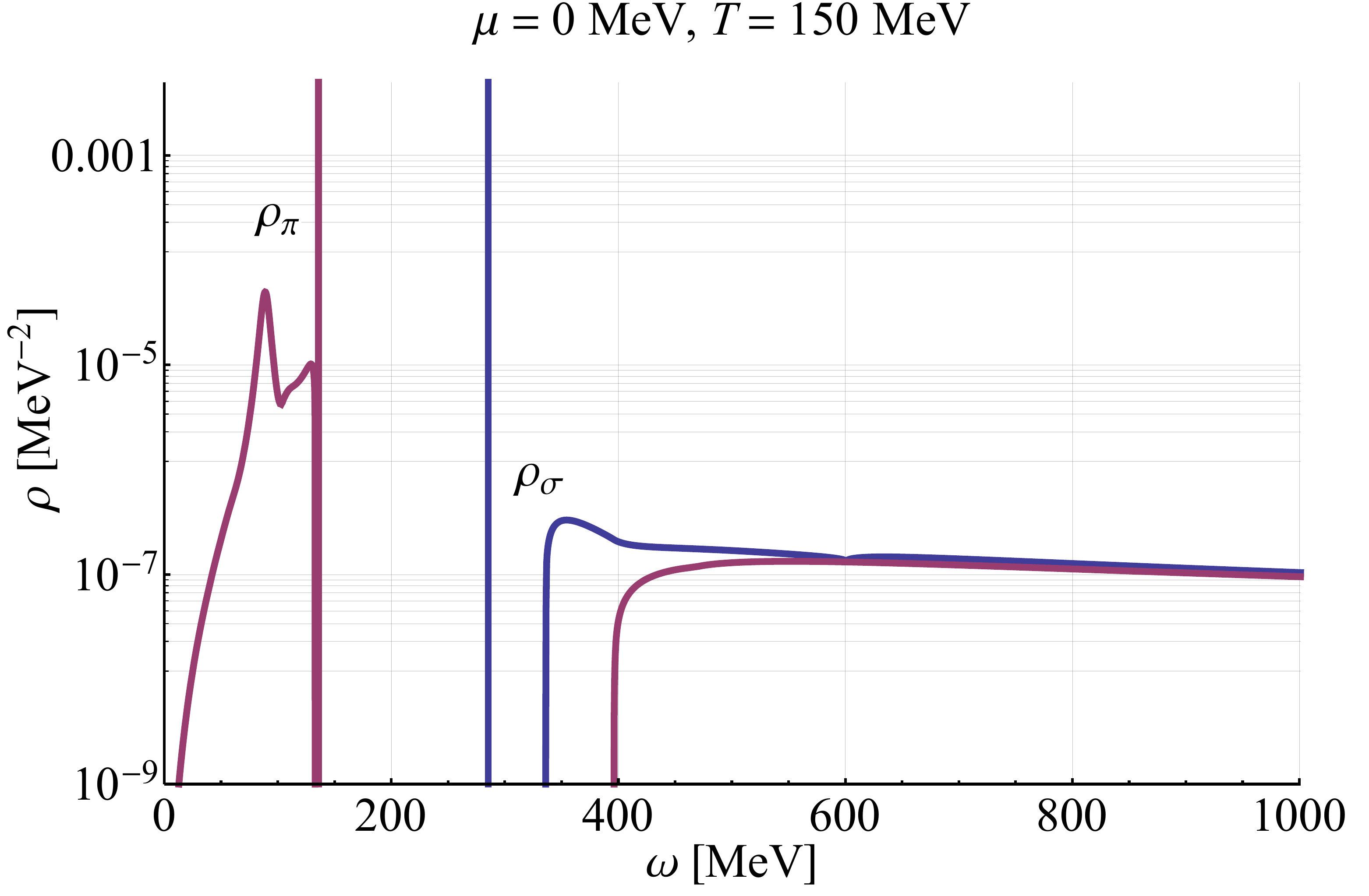}\hspace{2mm}
	\includegraphics[width=0.5\textwidth]{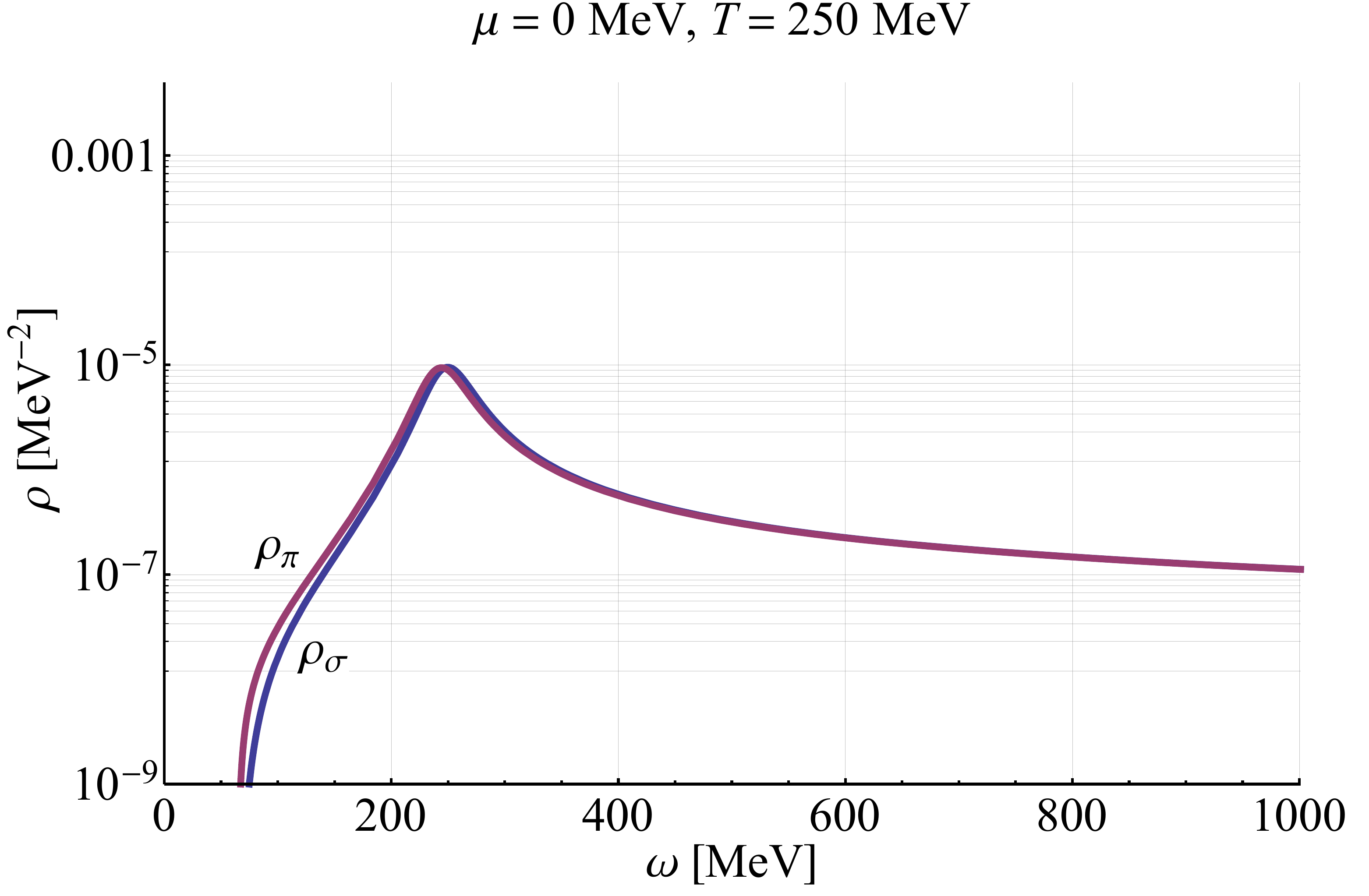}
	\caption{The sigma and pion spectral functions are shown for different temperatures at $\mu=0$ in the IR. The various decay channels give rise to a complicated structure of the in-medium spectral functions, see text for details. At high temperatures, the spectral functions become degenerate due to the restoration of chiral symmetry. Figure taken from \cite{Tripolt:2016cey}.}
	\label{fig:spectral_functions} 
\end{figure}

The flow equations for the mesonic two-point functions of the quark-meson model are shown diagrammatically in Fig.~\ref{fig:flow_equations_gamma2}. In the LPA the scale-dependent mesonic three- and four-point vertices are extracted from the effective potential while the quark-meson vertices are basically given by the Yukawa coupling $h$, for explicit expressions see \cite{Tripolt:2013jra}. The flow equations are analytically continued as described in Sec.~\ref{sec:analytic_continuation} and solved numerically. The resulting vacuum spectral functions for the pion and the sigma meson are shown in Fig.~\ref{fig:spectral_UV_IR}. At the UV scale, where chiral symmetry is unbroken (except for a small finite current-quark mass, i.e. $c\neq0$ in Eq. (4.1)) both spectral functions describe (nearly degenerate) massive stable particles. During the RG-evolution they are significantly modified by spontaneous chiral symmetry breaking. In the IR, the pion turns into a (pseudo)Goldstone mode and the sigma becomes unstable to decay into two pions, as it should be. The pion spectral function shows a continuum starting at the quark-antiquark threshold, $\omega\approx 600$~MeV. This threshold (also appearing in the sigma-channel) is an artifact of the non-confining nature of the QM model.

Fig.~\ref{fig:spectral_functions} shows the temperature dependence of the sigma and the pion spectral function at $\mu=0$. At $T=110$~MeV, the pion shows interesting modifications at lower energies due to processes where a pion captures another pion from the heat bath and turns into a sigma meson. In the vicinity of the chiral crossover, $T=150$~MeV, both the sigma and the pion are stable again while at $T=250$~MeV the spectral functions become degenerate as expected from the restoration of chiral symmetry.

\begin{figure}[t]
	\includegraphics[width=0.49\textwidth]{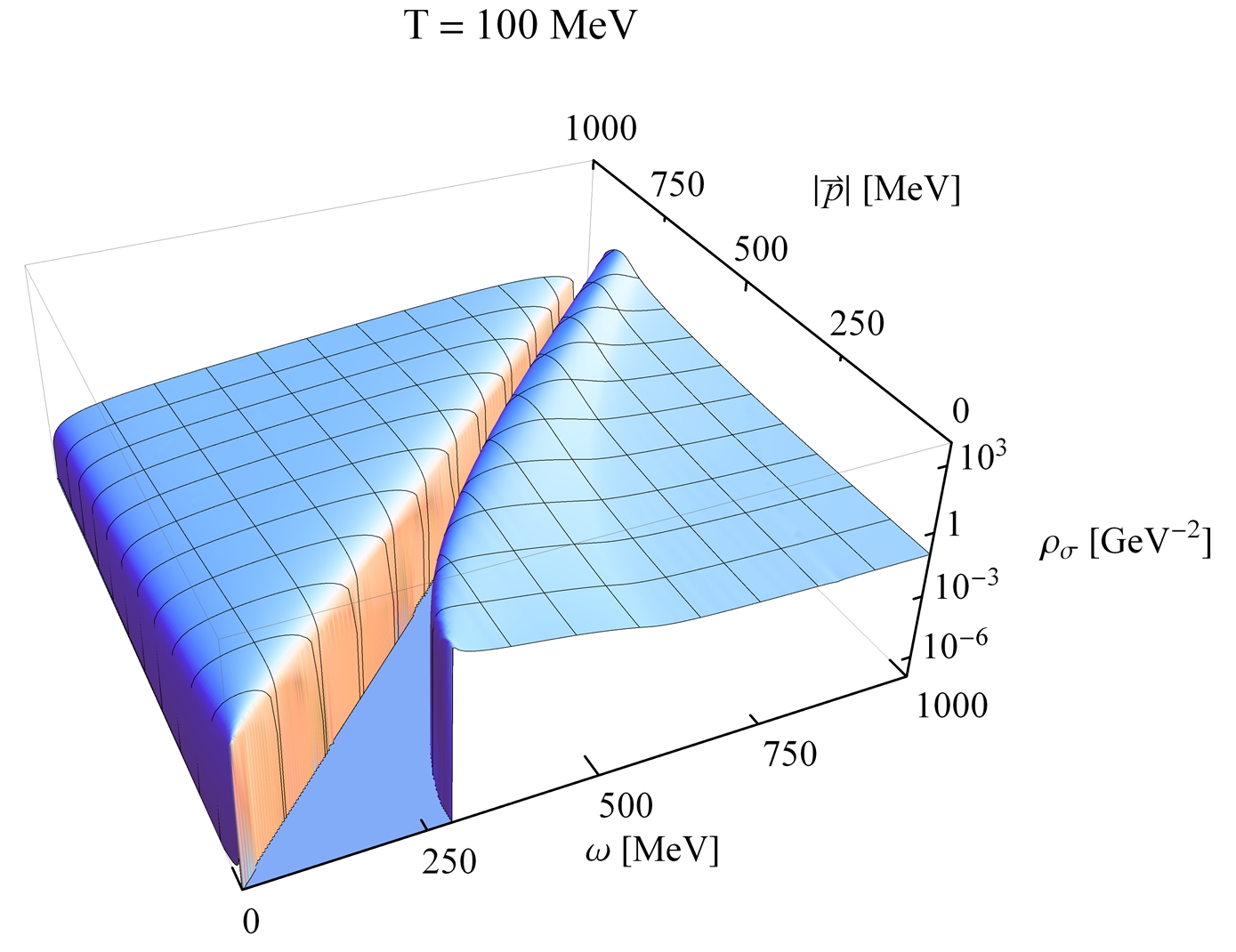}
	\includegraphics[width=0.49\textwidth]{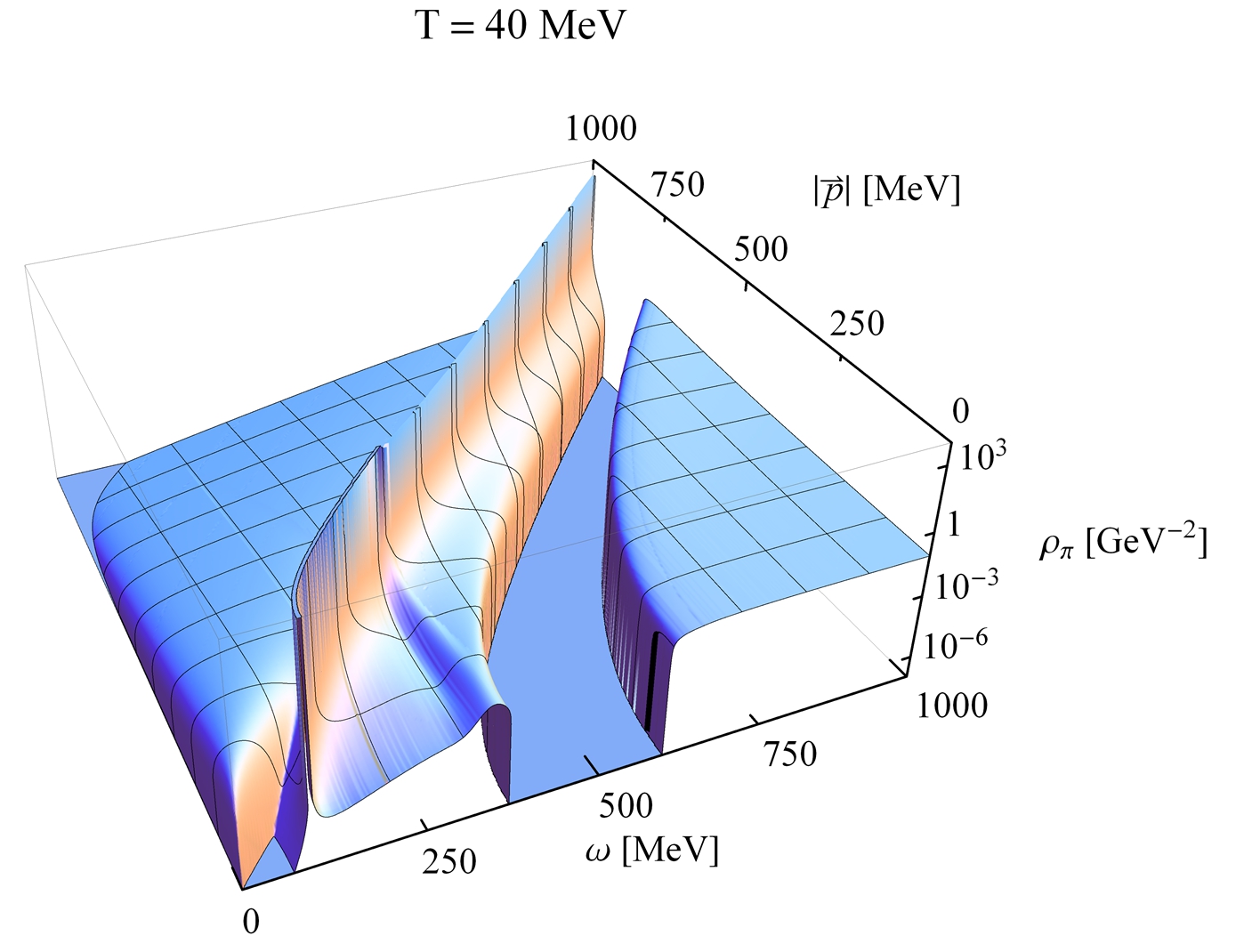}
	\caption{Left: The sigma spectral function is shown vs.~energy $\omega$ and spatial momentum $|\vec{p}|$ at $T=100$~MeV and $\mu=0$. For $|\vec{p}|>\omega$, the spectral function is generated by space-like processes while the time-like region, $|\vec{p}|<\omega$, is Lorentz-boosted towards the light cone as $|\vec{p}|$ increases. Right: Same as left but for the pion spectral function at $T=40$~MeV. Figure taken from \cite{Tripolt:2016cey}.}
	\label{fig:spectral_3D} 
\end{figure}

Fig.~\ref{fig:spectral_3D} shows the dependence of the sigma and the pion spectral function on the spatial momentum at selected temperatures and $\mu=0$. The space-like regime, $|\vec{p}|>\omega$, is generated by processes where a space-like particle excitation is absorbed by the heat bath. The time-like regime, where $|\vec{p}|<\omega$, is Lorentz-boosted toward the light cone as the spatial momentum increases. As a result of thermal capture processes the pion spectral function in addition shows structures with a nearly flat dispersion relation, see \cite{Tripolt:2014wra, Jung:2016yxl} for details.

\section{Vector- and Axial-Vector-Meson Spectral Functions}
\label{sec:vector}

We now turn to the in-medium properties of vector mesons, which are of relevance to dilepton signals in heavy-ion collisions. We will use the extended linear-sigma model introduced in \cite{Jung:2016yxl} where the $\rho$ vector meson and the $a_1$ axial-vector meson are introduced as gauge bosons of a (local) chiral symmetry. The effective action of this model reads:
\begin{align}
\label{eq:effective_action_vector}
\Gamma_k = &\int d^4x \Big[\bar{\psi} \left(\slashed{\partial}-\mu\gamma_0+
h_S\left(\sigma +\mathrm{i} \vec{\tau}\vec{\pi}\gamma_5\right)+
\mathrm{i} h_V \left(\gamma_{\mu} \vec{\tau}\vec{\rho}^{\mu}+\gamma_{\mu}\gamma_5 \vec{\tau}\vec{a}_1^{\mu}\right)
\right)\psi+ U_{k}(\phi^2)-c\sigma+\frac{1}{2} (\partial_{\mu}\phi)^2 \nonumber\\
&+ \frac{1}{8} \rm{Tr}\left(\partial_{\mu}V_{\nu}-\partial_{\nu}V_{\mu}\right)^2-
\mathrm{i}g V_{\mu}\phi\partial_{\mu}\phi-\frac{1}{2}g^2\left(V_{\mu}\phi\right)^2+\frac{1}{4}m_{k,V}^2 \rm{Tr}\left(V_{\mu}V_{\mu}\right)
\Big] +\Delta\Gamma_{\pi a_1}\,,
\end{align}
where the vector mesons are given in the adjoint representation of $O(4)$ with
\begin{align}
V_\mu = \vec{\rho}^{\,\mu}\vec{T}+\vec{a}_1^{\,\mu}\vec{T}^5,
\end{align}
and the term $\Delta\Gamma_{\pi a_1}$, which is connected to $\pi-a_1$ mixing, is given by
\begin{align}
\Delta\Gamma_{\pi a_1} &= \int_x \!\bigg\{ g\, \sigma_0\, \vec{a}_1^{\,\mu}\!\cdot \partial_\mu\vec{\pi} - \frac{1}{2} \frac{g^2 \sigma_0^2}{m_{k,V}^2 + g^2 \sigma_0^2} (\partial_\mu\vec{\pi})^2
-\frac{g^2 \sigma_0^2}{m_{k,V}^2 + g^2 \sigma_0^2}\, \vec{\rho}^{\,\mu}\!\!\times\!\vec{\pi}\cdot\partial_\mu\vec{\pi} \bigg\}\,.
\end{align} 
In the UV, the effective potential is chosen to be of the form
\begin{align}
U_{\Lambda} = b_1 \phi^2+b_2
\phi^4\, ,
\end{align}
where explicit values for the parameters are listed in Tab.~\ref{tab:parameters2}. In the IR we obtain the following vacuum values for the pion decay constant and the Euclidean masses: $f_{\pi}\equiv\sigma_0=93.0$~MeV, $m_{\pi}=140$~MeV, $m_{\sigma}= 557$~MeV and $m_{\psi}=300$~MeV. The gauge coupling $g$ and the vector-mass parameter $m^2_{k,V}$ have been chosen such as to obtain a $\rho$ pole mass of $789$~MeV and an $a_1$ pole mass of $1275$~MeV.

\begin{figure}[t]
	\includegraphics[width=0.99\textwidth]{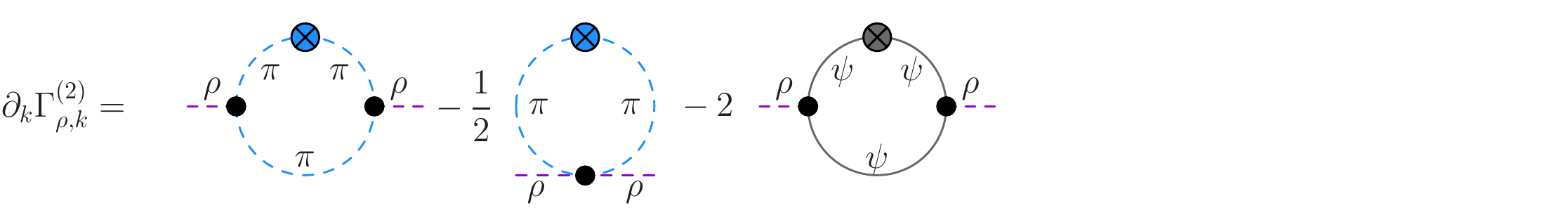}\\
	\includegraphics[width=0.99\textwidth]{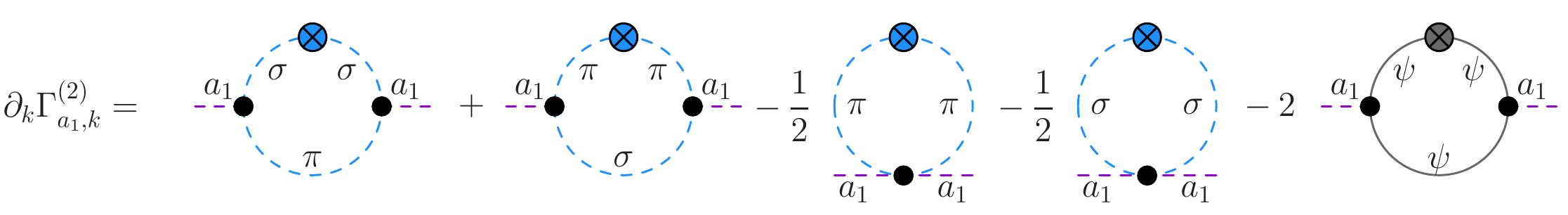}
	\caption{Diagrammatic representation of the flow equations for the $\rho$ and $a_1$ two-point functions. Dashed lines denote mesonic propagators, solid lines fermionic propagators, crossed circles correspond to regulator insertions, $\partial_k R_k$, while black points denote three- and four-point vertices. Figure taken from \cite{Jung:2016yxl}.}
	\label{fig:flow_equations_vector} 
\end{figure}

\begin{table}[b]
	\centering
	\begin{tabular}{C{1.6cm}|C{0.6cm}|C{1.7cm}|C{1.4cm}|C{0.6cm}|C{1.7cm}}
		$b_1 \text{~[MeV}^2\text{]}$ & $b_2$ & $c\text{~[MeV}^3\text{]}$ & $h_S=h_V$&$g$ & $m_{\Lambda,V}\text{~[MeV}\text{]}$\\
		\hline\hline
		857300  & 0.2 & 1.8228$\cdot 10^6$ &3.226 & 11.4 & 1450
	\end{tabular}
	\caption{Parameter set used for the gauged linear-sigma model. The UV cutoff is chosen to be $\Lambda = 1500$~MeV.}
	\label{tab:parameters2} 
\end{table}

The diagrammatic structure of the flow equations of the $\rho$ and $a_1$ two-point functions is shown in Fig.~\ref{fig:flow_equations_vector}. We note that so far loops with dynamical (axial-)vector mesons have been neglected. These flow equations have components that are transverse and longitudinal to the heat bath which can be separated by using the following projection operators,
\begin{align}
\label{eq:projection_heat-bath}
\begin{split}
\Pi^{T,\perp}_{\mu\nu}(p) &= \begin{cases} 0 & \text{if}\; \mu=0\; \text{or}\; \nu=0\\ \delta_{\mu\nu}-\frac{p_\mu p_\nu}{\vec{p}^2} & \text{else} \end{cases}\,,\\
\Pi^{T,\parallel}_{\mu\nu}(p) &= \delta_{\mu\nu}-\frac{p_\mu p_\nu}{p^2} - \Pi^T_{\mu\nu}(p)\,,
\end{split}
\end{align}
with
\begin{align}
\Pi^{T}_{\mu\nu}(p) = \Pi^{T,\perp}_{\mu\nu}(p)+\Pi^{T,\parallel}_{\mu\nu}(p) = \delta_{\mu\nu}-\frac{p_{\mu}p_\nu}{p^2}.
\end{align}
For further details on the theoretical setup and the numerical implementation we refer to \cite{Jung:2016yxl}.

\begin{figure}[t!]
	\includegraphics[width=0.48\textwidth]{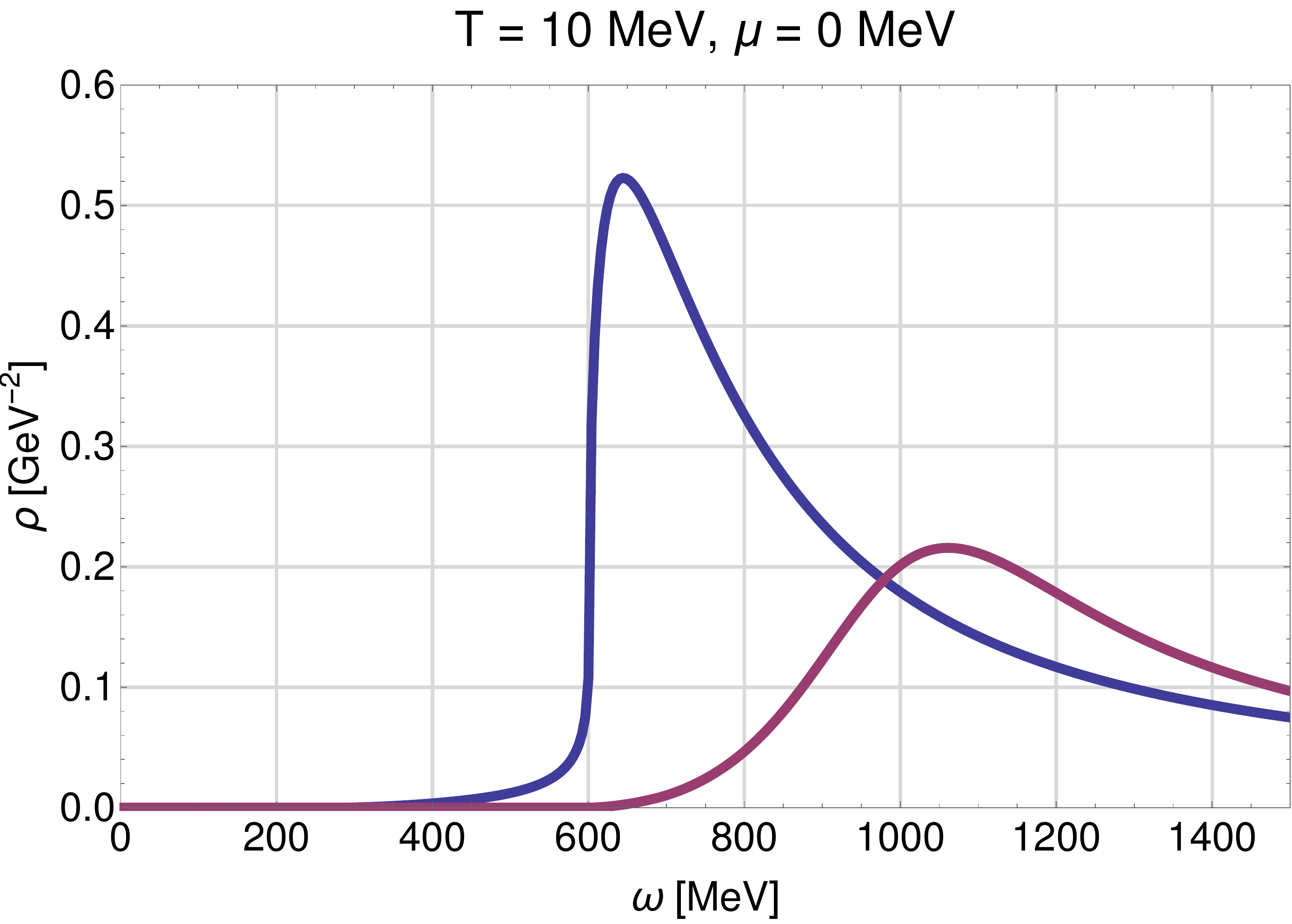}\hspace{5mm}
	\includegraphics[width=0.48\textwidth]{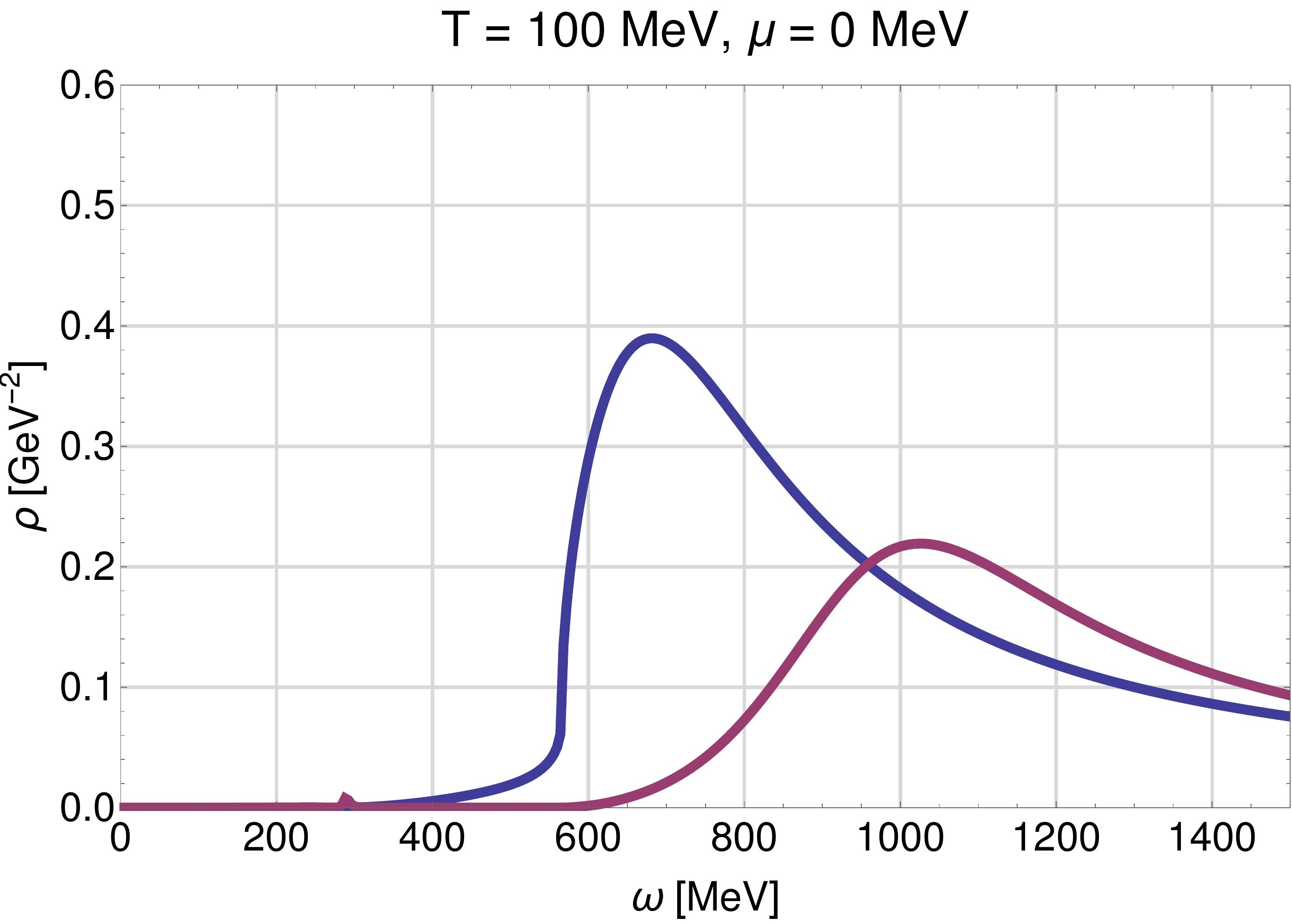}\\[2mm]
	\includegraphics[width=0.48\textwidth]{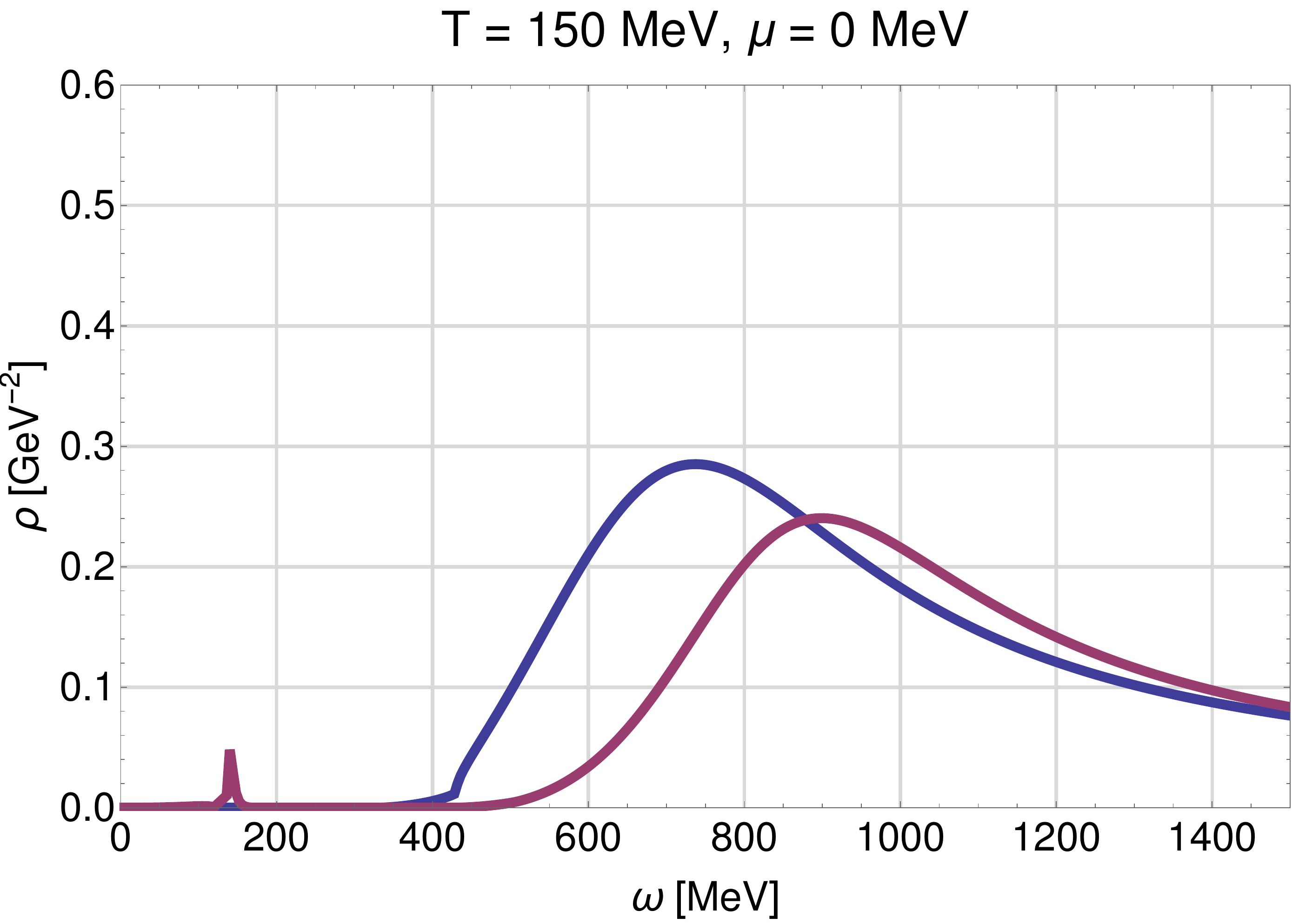}\hspace{5mm}
	\includegraphics[width=0.48\textwidth]{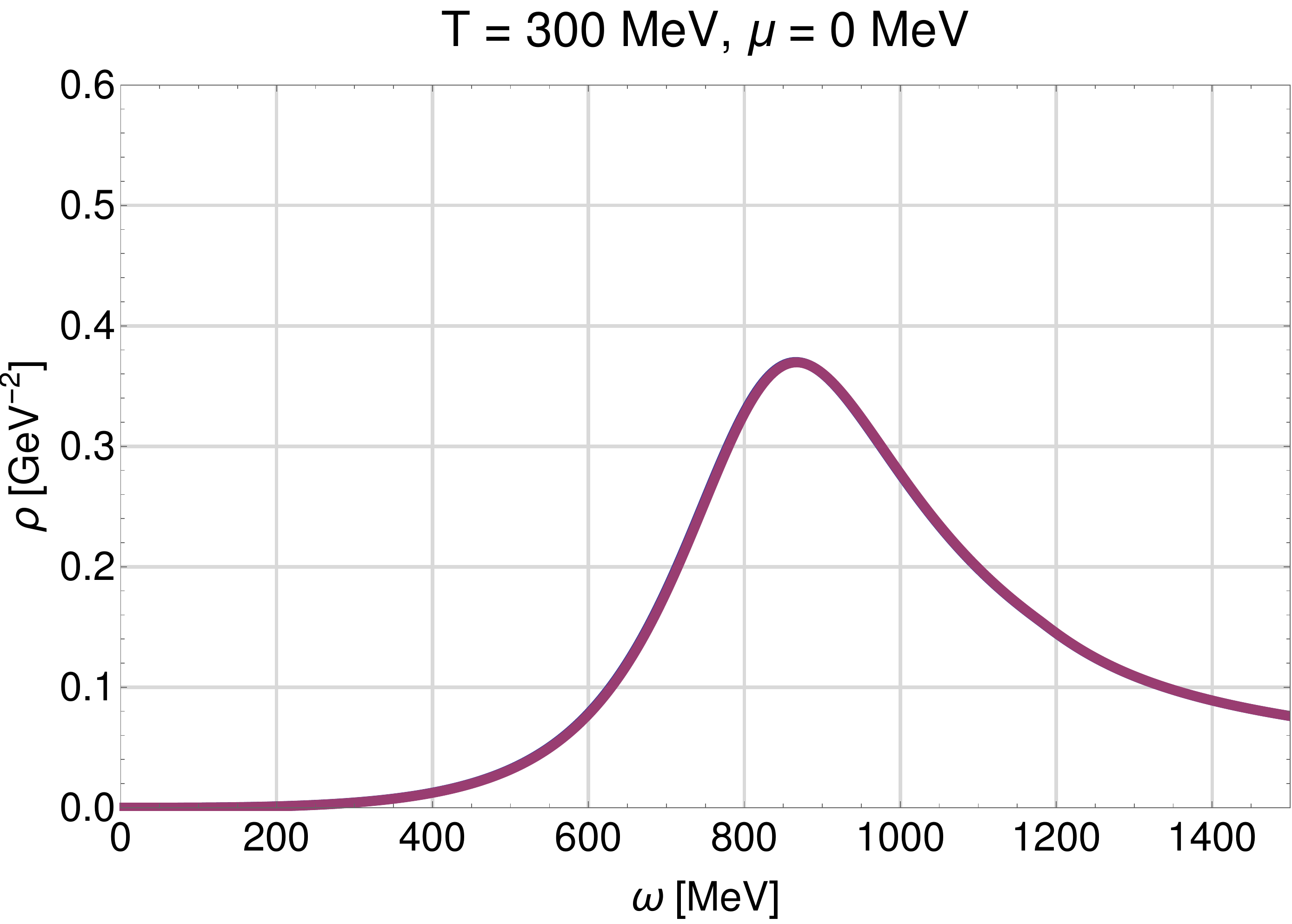}
	\caption{The $\rho$ and $a_1$ spectral functions are shown for different temperatures at $\mu=0$ in the IR. The $\rho$ spectral function is dominated by the decay channels $\rho\rightarrow \pi+\pi$ and $\rho\rightarrow \psi+\bar\psi$ while the $a_1$ spectral function is mostly affected by the decay channels $a_1\rightarrow \sigma+\pi$ and $a_1\rightarrow \psi+\bar\psi$, see text for details. At high temperatures, the spectral functions become degenerate due to the restoration of chiral symmetry.}
	\label{fig:vector_spectral_functions} 
\end{figure}

\begin{figure}[t]
	\centering\includegraphics[width=0.57\textwidth]{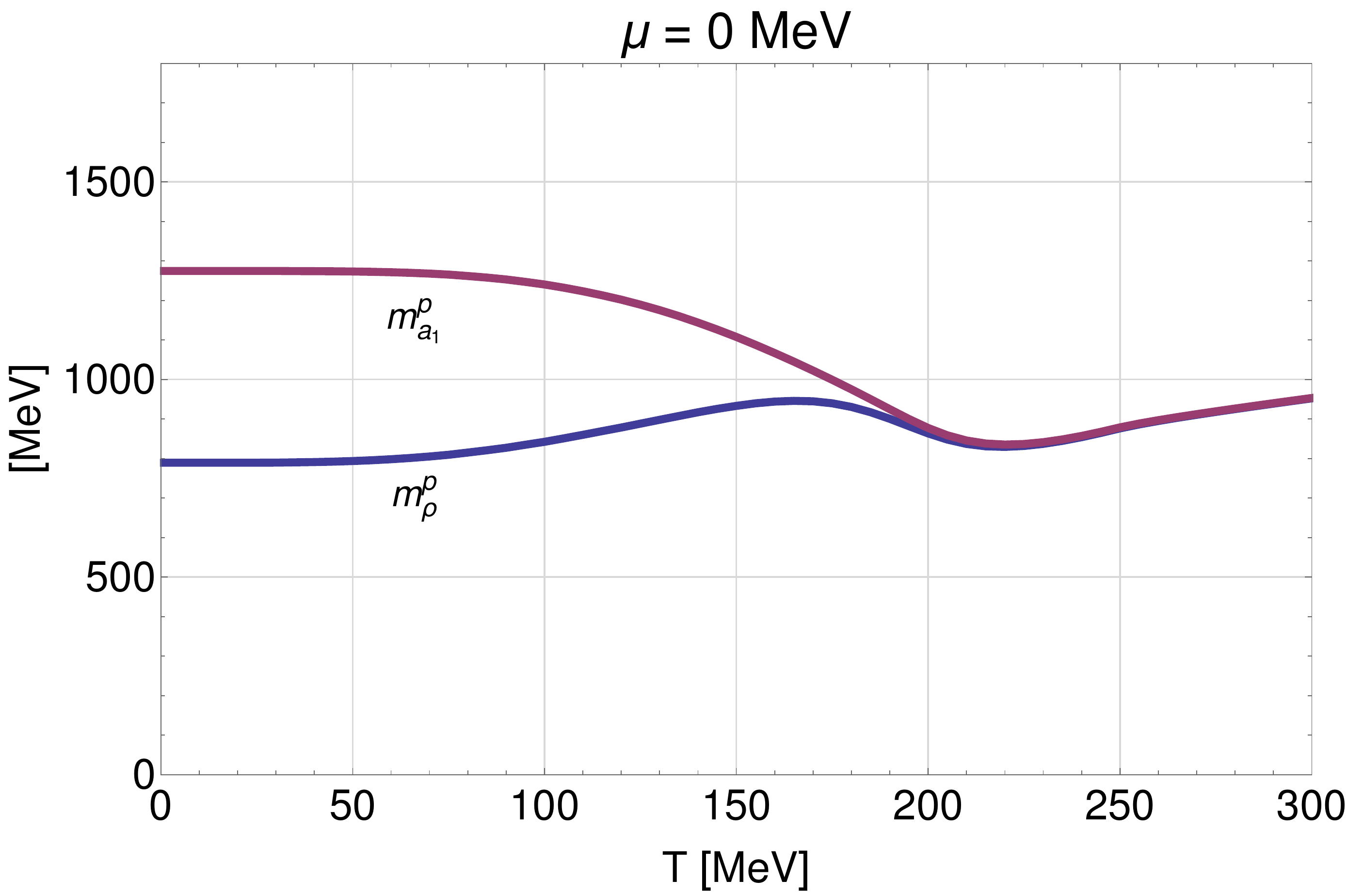}\\
	\caption{The pole masses of the $\rho$ and the $a_1$ mesons are shown as a function of temperature at $\mu=0$. The $\rho$ pole-mass slightly increases towards the chiral crossover while the $a_1$ pole-mass drops down, which is consistent with a `broadening-$\rho$ scenario'. The pole masses become degenerate at $T\approx200$~MeV due to the restoration of chiral symmetry. Figure taken from \cite{Jung:2016yxl}.}
	\label{fig:pole_masses} 
\end{figure}

Results for the transverse components of the $\rho$ and the $a_1$ spectral function are shown in Fig.~\ref{fig:vector_spectral_functions}. Since also this model does not incorporate confinement, the spectral functions, already in the vacuum, are strongly influenced by the quark-antiquark decay channel which opens up at $\omega\approx600$~MeV. Nevertheless, the $\rho$ shows a characteristic peak at $\omega\approx700$~MeV, which is also influenced by the $\rho\rightarrow \pi+\pi$ channel, while the $a_1$ peak is much broader. With increasing temperature the $\rho$ peak melts but essentially remains at the same energy while the broad $a_1$ resonance moves towards lower energies. For temperatures above the crossover transition, the spectral functions again become degenerate due to the restoration of chiral symmetry. The evolution of the pole masses with increasing temperature is also shown in Fig.~\ref{fig:pole_masses}. While the $\rho$ mass slightly increases towards the chiral crossover, the $a_1$ mass becomes smaller and approaches the $\rho$ mass at $T\approx 200$~MeV.

\section{Summary}
\label{sec:summary}

In these proceedings we have summarized recent results on in-medium spectral functions of hadrons using the Functional Renormalization Group method. Based on a recently developed analytic continuation scheme, we have discussed how the flow equations for the inverse propagators can be solved directly for real energies. 

The quark-meson model has been used to obtain the in-medium spectral functions for the pion and its chiral partner, the sigma meson. Due to various additional decay channels and thermal processes, they show a complicated structure, including a space-like response. It could be shown that the spectral functions become degenerate for all energies and three-momenta at high temperatures beyond the chiral crossover as expected from restoration of chiral symmetry on very general grounds.

To study the in-medium properties of vector and axial-vector mesons, a gauged linear-sigma model including quarks has been used. Results for the spectral functions of the $\rho$ and the $a_1$ at finite temperature have been presented and their degeneration at high temperatures was observed. Moreover, the behavior of their pole masses is consistent with a `broadening-$\rho$' scenario, obtained from more phenomenological approaches and supported by experiment \cite{Rapp:1999ej,Damjanovic:2007qm}.  

Future improvements of the presented framework, in particular concerning a realistic description of (axial-)vector meson spectral functions, will involve the inclusion of wave-function renormalization factors and of decay channels into vector mesons. Also the extension towards baryonic degrees of freedom and the calculation of dilepton rates represent exciting applications.

\section*{Acknowledgments}

This work was supported by the Deutsche Forschungsgemeinschaft (DFG) through the grant CRC-TR 211 ``Strong-interaction matter under extreme conditions,'' and by the German Federal Ministry for Education and Research (BMBF) through grants 05P12VHCTG and 05P16RDFC1. F.~R.~also acknowledges support by the Austrian FWF through grant P24780-N27, the DFG Collaborative Research Centre SFB-1225 ``ISOQUANT,'' and by the DFG grant RE 4174/1-1.

\end{document}